\documentclass[]{spie}  

\usepackage{amsmath,amsfonts,amssymb}
\usepackage{graphicx}
\usepackage[colorlinks=true, allcolors=blue]{hyperref}
\usepackage{upgreek}
\usepackage{comment}

\title{Scientific modeling of Optical 3D Measuring Devices based on GPU-accelerated Ray Tracing using the NVIDIA OptiX Engine}

\author{Keksel A.}
\author{Schmidt S.}
\author{Beck D.}
\author{Seewig J.}
\affil{University of Kaiserslautern-Landau, Kaiserslautern, Germany}

\authorinfo{Further author information: (Send correspondence to A. Keksel)\\A. Keksel: E-mail: andrej.keksel@mv.rptu.de, Telephone: +49 631 205 4439}

\begin{document} 
\maketitle

\begin{abstract}
Scientific optical 3D modeling requires the possibility to implement highly flexible and customizable mathematical models as well as high computing power. However, established ray tracing software for optical design and modeling purposes often has limitations in terms of access to underlying mathematical models and the possibility of accelerating the mostly CPU-based computation. To address these limitations, we propose the use of NVIDIA's OptiX Ray Tracing Engine as a highly flexible and high-performing alternative. OptiX offers a highly customizable ray tracing framework with onboard GPU support for parallel computing, as well as access to optimized ray tracing algorithms for accelerated computation. To demonstrate the capabilities of our approach, a realistic focus variation instrument is modeled, describing optical instrument components (light sources, lenses, detector, etc.) as well as the measuring sample surface mathematically or as meshed files. Using this focus variation instrument model, exemplary virtual measurements of arbitrary and standardized sample surfaces are carried out, generating image stacks of more than 100 images and tracing more than 1E9 light rays per image. The performance and accuracy of the simulations are qualitatively evaluated, and virtually generated detector images are compared with images acquired by a respective physical measuring device.
\end{abstract}


\section{INTRODUCTION}
Optical measuring devices are widely used in surface metrology because they provide a fast, high resolution, non-contact method of measurement. The aforementioned characteristics of optical topography measurement systems generally make these systems very attractive for industrial applications, especially considering measurement tasks where tactile methods cannot be applied in a reasonable way due to either limited time resources or poor accessibility of the feature to be characterized. However, the calibration of optical measuring instruments is more challenging than the calibration of tactile instruments. This is due to the complexity of the interaction between the sample surface and the optical measuring system. The quality of the achieved (or achievable) measurement results strongly depends on the entire surface structure (shape, roughness, inclination) as well as the material of the measuring sample. Under difficult measuring conditions, this can lead to significant deviations between the measured surface structure and the actual underlying surface structure, thus severely restricting the general traceability of optical measurements to optically cooperative surfaces \cite{Leach.2011}.

One way of ensuring the traceability of optical measurements is to set up numerical models to provide a priori estimation of system responses when measuring surfaces of any complexity, in order to estimate expected measurement deviations in advance of the actual measurement. These models, so-called virtual measurement systems, are characterized by the fact that they take into account various measurement influencing factors of relevance and thus reproduce the real measurement process comprehensively in the context of a precise physical model \cite{Schmitt.2008, Groot.2004, Seewig.2016}. Underlying influencing factors can be varied quickly and on a large scale on the basis of suitable physical models. The model-based prediction of measurement results can thus be applied, in addition to the a priori estimation of expected measurement deviations, e.g. to increase the fundamental understanding of physical measurement processes and thus to optimize existing measurement systems or even to provide new insights for the development of new measurement systems \cite{Schmitt.2008}.

Ray tracing (RT), which has always been computationally intensive, has long been a commonly used approach for modeling optical devices and processes. As computing power has increased over the years, so have the requirements for RT simulations (in particular the number of rays to be traced for sufficiently good simulations), so that the simulation time is still the limiting factor for the widespread use of RT approaches in the context of optical 3D modeling. Current developments in the field of RT, which mainly originate from the visualization of photorealistic scenes for movies and video games, tend to parallelize the intersection computations required for RT by offloading them to the graphics processing unit (GPU), which accelerates them enormously and reduces the required simulation times accordingly \cite{Mauch.2013}.
In this paper the fundamentals of a highly flexible ray tracing based physical model of a focus variation (FV) instrument for the acquisition of areal surface topographies are presented.

\section{NVIDIA OPTIX RAY TRACING ENGINE}
The NVIDIA OptiX Ray Tracing Engine is a powerful and highly flexible development environment for implementing sophisticated RT applications. OptiX was designed on the premise that most RT algorithms can be implemented using a small number of programmable operations, regardless of the specific application. The main application areas of the OptiX engine are games, movies, and commercials, where it is used for the (real-time) generation of impressive and highly realistic looking three-dimensional scenes \cite{Parker.2010, Parker.2013}. In contrast, the application of OptiX in the field of scientific optical modeling is not yet widespread, although the potential of OptiX in this context has already been demonstrated \cite{Mauch.2013, Felbecker.2012, Blyth.2017, Blyth.2019, Blyth.2021}. Briefly, the key aspects that make OptiX attractive for scientific modeling are as follows.\\

GPU-accelerated parallel computation and access to powerful RT algorithms:\\
RT applications are considered to be computationally intensive, as generating reliable results usually requires tracing a very large number of independent rays and taking into account a large number of ray interactions. Using OptiX, these calculations are offloaded to the graphics card unit (GPU). This has the great advantage that a large number of calculations are executed in parallel, speeding up the simulation considerably (up to a factor of 100$\times$, compared to CPU computation \cite{Mauch.2013}). In addition, the OptiX engine provides access to highly specialized and optimized RT algorithms. The only prerequisite is the use of a supported NVIDIA GPU \cite{NVIDIACorporation.RT}.\\

High level of flexibility:\\
OptiX is designed to allow highly optimized RT algorithms to be adapted to user-defined requirements and boundary conditions of optical simulations. Accordingly, no physical laws of optics (such as reflection laws, refraction laws/models, absorption and scattering models, etc.) are predefined. The optical models the simulation is based on can - and must - be defined by the user without any restrictive pre-defined specifications. This may sound like a disadvantage at first, but in the context of scientific optical modeling it turns out to be essential with respect to often very complex model assumptions and newly developed models.
The user benefits from highly specialized and efficient RT algorithms as well as from maximum flexibility in model development and design. As a result, scientific work can focus on the development and implementation of physical models without sacrificing high efficiency in the execution of usually very complex RT simulations.
The description of the features and capabilities of OptiX has been heavily abbreviated to fit the context of scientific optical modeling. Detailed information and technical details can be found in the documentation provided by NVIDIA \cite{NVIDIACorporation.Documentation}.

\section{VIRTUAL FOCUS VARIATION INSTRUMENT}
Within this paper, the potential of GPU-accelerated RT using OptiX for scientific optical modeling and simulation is demonstrated using the example of the implementation of a virtual focus variation (FV) instrument. In addition to the implementation of volumetric objects such as measuring instrument components (lenses, mirrors, etc.) and samples to be measured virtually (volumetric bodies or three-dimensional surface topographies), the physical behavior of light-matter interaction (refraction, reflection, scattering, etc.) has to be defined and implemented. The various steps for the implementation of a comprehensive virtual FV instrument within the framework of the OptiX development environment are described in detail below.

\subsection{Measuring device components}
The basic structure of the optical system of a FV instrument is similar to that of a digital microscope. The implementation of the hardware components of the measuring device relevant for the model is limited in the context of this work to light source, lenses, mirrors, beam splitter and detector installed in the measuring device. The light source is modeled as a flat circular light emitting surface. The ideal intensity-sensitive detector (representing a CCD sensor) is represented by a square planar surface with defined geometric dimensions and defined pixel density. The optical system of the modeled digital microscope, i.e. the geometric dimensions of the modeled lenses, beam splitters and mirrors and their relative arrangement, is provided by Bruker Alicona. For confidentiality reasons, the design and implementation details of the optical system are not described here.
\subsection{Measuring sample}
The geometry of the virtual measuring samples is reduced to their surface topography. In addition to idealized test structures such as (inclined) planes, steps, spherical surfaces, and sinusoidal and chirp structures, irregular (standardized) surfaces whose structures are based on those of real technical surfaces can also be implemented. The surfaces can be implemented as mesh (based on a discrete point cloud) or as continuously described function. This is also valid for the implementation of hardware components like lenses. Simple structures that can be described continuously (implicitly) are e.g. spherical surfaces, while triangulated descriptions, e.g. point clouds meshed by Delaunay triangulation and exported in the widely used STL file format, are well suited for irregular surfaces.
\subsection{Light-matter interaction}
The light source to be modeled is defined not only by its geometry, but also by its radiation characteristics. A common model to describe the radiation characteristics of technical light sources is Lambert's cosine law. Light sources with near Lambertian radiation characteristics are characterized by the fact that they appear uniformly bright from any viewing angle \cite{Smith.2008}.
In addition, the interactions of light with the optical system must be described. Ideal assumptions are made regarding light reflection and refraction at transitions between materials of different optical densities (here, for example, the transition from air to lens glass and vice versa). These assumptions are based on Snell's law of refraction.
The interaction of light rays with the measuring sample is assumed to be opaque and non-absorbing. In this simple model, the physical interaction of light with the measuring sample is reduced to the scattered reflection of the light at the sample's surface.
A simple model for describing light scattering at the sample surface is the Henyey-Greenstein scattering function, which is adapted to surface scattering to be modeled within this paper. This function was originally developed to describe the scattering of light by interstellar dust clouds \cite{Henyey.1941}. It was later shown to be suitable for approximating the angular scattering of single scattering events in biological tissues \cite{Tuchin.1993}. The main feature of the Henyey-Greenstein scattering function is that the scattering characteristics can be continuously adjusted from Lambertian scattering ($g\rightarrow 0$) to specular reflection ($g\rightarrow 1$) by adjusting a single parameter ($g$).
\subsection{Scanning process}
To convert a digital microscope into an FV instrument for the acquisition of 3D topographies, it must be equipped with a precise vertical translation unit, e.g. a piezo motor with step sizes in the nanometer range. In practice, for measurements of structures in the sub-µm to several hundred µm range, image stacks with several hundred to thousand images per measurement are acquired. The virtual FV instrument must be able to handle this volume in a reasonable time as well.
Virtual microscopy using RT requires the tracing of more than 1E9 rays per image, depending on the desired quality of the results. For the simulation of single images taken by the instrument, simulation times of a few minutes up to a few hours are still feasible. However, this is no longer the case when simulating a stack of images consisting of hundreds or thousands of single images. In this case, the acceleration of the computations is crucial. This is done by default in OptiX by parallelizing the calculations by offloading them to the GPU and simultaneously using highly specialized RT algorithms.

\section{VIRTUAL MEASUREMENTS}
\subsection{Virtual imaging of a rough plateau structure}
\begin{figure} [ht]
   \begin{center}
   \begin{tabular}{c} 
   \includegraphics[height=8cm]{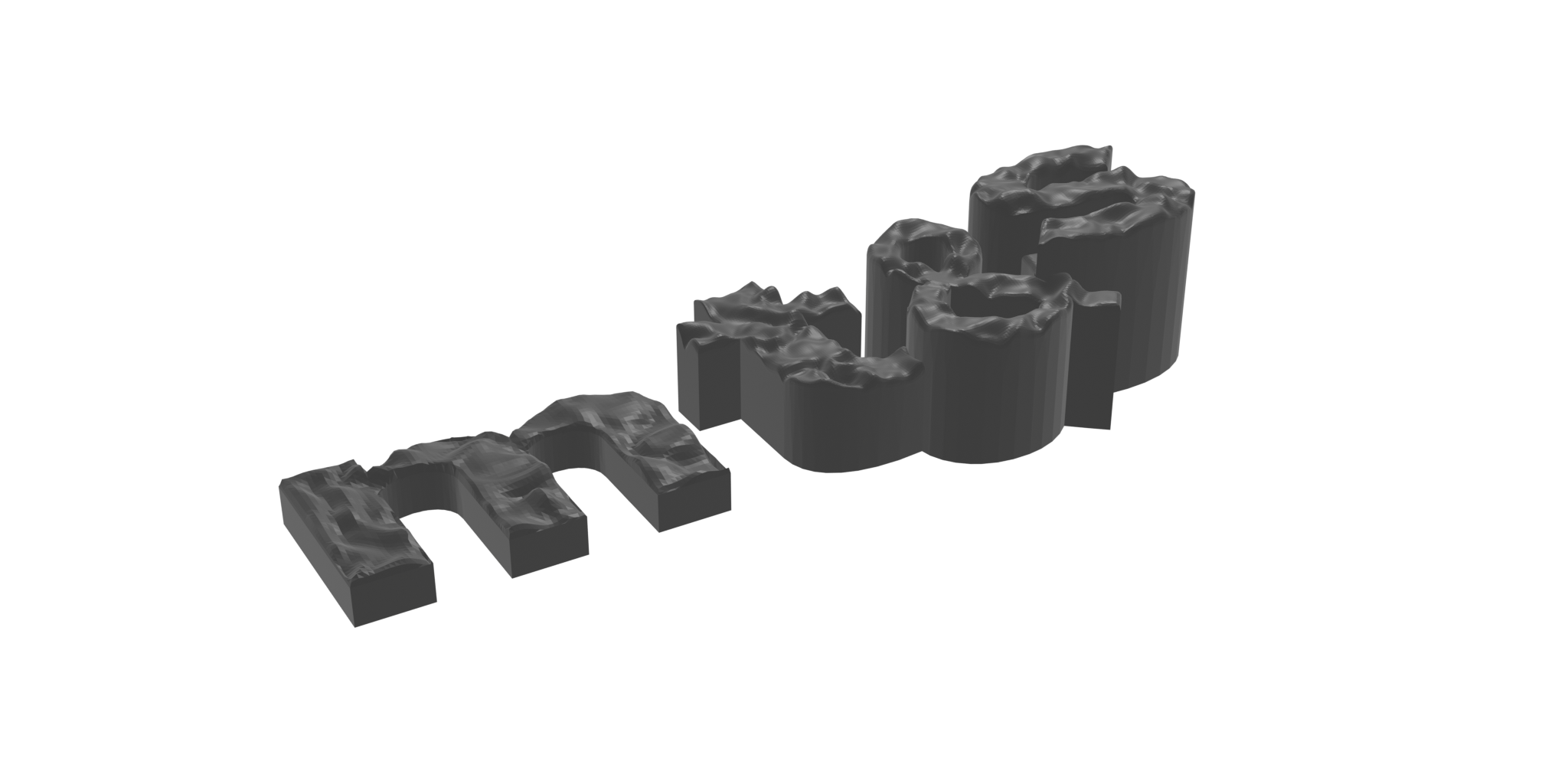}
   \end{tabular}
   \end{center}
   \caption[example] 
   { \label{fig:1_mts_logo_3D_model} 
3D model of the virtual measuring sample containing irregular surface structures.}
   \end{figure}

To demonstrate the quality of the results and the performance of the virtual FV instrument, exemplary detector images simulated under different conditions are visualized and compared with each other. The virtual measuring sample consists of plateaus at different height levels containing irregular surface structures (see 3D model in Fig. \ref{fig:1_mts_logo_3D_model}). 

\begin{figure} [ht]
   \begin{center}
   \begin{tabular}{c} 
   \includegraphics[width=17cm]{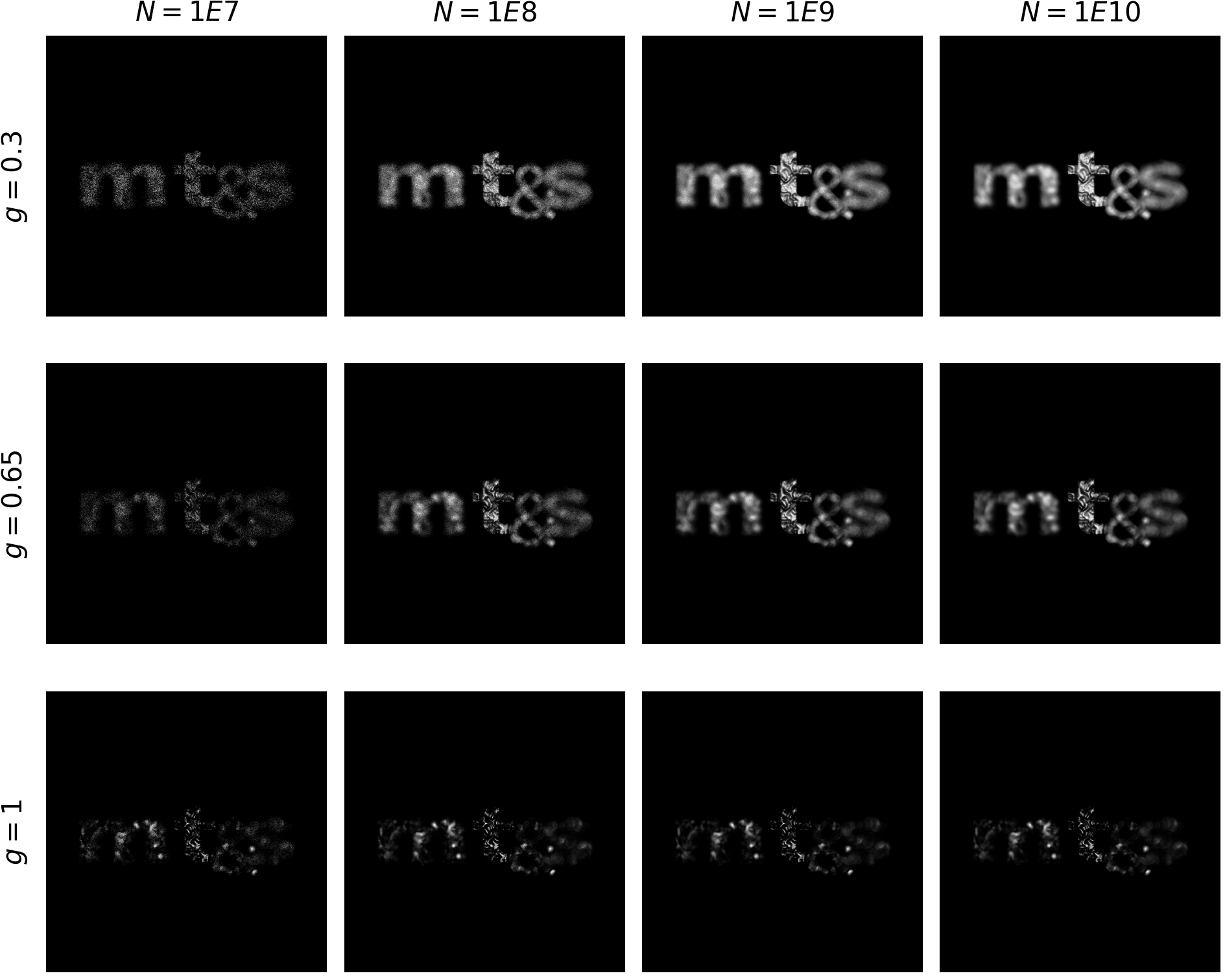}
   \end{tabular}
   \end{center}
   \caption[example] 
   { \label{fig:2_mts_logo_parameter_variation_images} 
Comparison of simulated detector images varying the number of traced rays (increasing from left to right) and varying the optical surface roughness (becoming optically smoother from top to bottom).}
   \end{figure}

In the context of the simulations performed here, an additional ``optical surface roughness" is assigned to the surface of the measuring sample in the format of a scattering angle distribution according to the aforementioned Henyey-Greenstein scattering function described above with $g\in[0,1]$, where $g=0$ corresponds to Lambertian scattering and $g=1$ to ideal specular reflection. This highly simplified scattering model can be interpreted as an additional stochastic micro-roughness on top of the deterministic surface structure.
The results shown in Fig. \ref{fig:2_mts_logo_parameter_variation_images} were generated under full factorial variation of the following simulation parameters:

\begin{table}[ht]
\caption{Simulation parameters of the full factorial parameter variation performed.} 
\label{tab:Logo_Simulation_Parameters}
\begin{center}       
\begin{tabular}{|l|l|} 
\hline
\rule[-1ex]{0pt}{3.5ex}  Total number of light rays traced $N_{\text{Rays}}$  & 1E7, 1E8, 1E9, 1E10 \\
\hline
\rule[-1ex]{0pt}{3.5ex}  Optical surface roughness $g$ & 0.3, 0.65, 1 \\
\hline
\end{tabular}
\end{center}
\end{table}
The surface of the letter ``t" of the ``mt\&s" logo is located in the focal plane of the lens when simulating the detector images shown in Fig. \ref{fig:2_mts_logo_parameter_variation_images} and using NVIDIA Titan V GPU Hardware, the simulation time to generate these detector images took, on average, $\sim$55 seconds per 1E9 rays.
As the number of light rays traced increases (from left to right in Fig. \ref{fig:2_mts_logo_parameter_variation_images}), a qualitative visual comparison indicates that the number of light rays required to obtain high-quality results with a low noise level should be chosen in the range of $\geq$1E9. A further increase in the number of rays does not lead to a significant increase in the quality of the results, at least not visually in this example. Conversely, when simulating with $<$1E9 rays, the quality of the results decreases visibly.

The statements made about the necessary number of rays to be traced are valid in this example for all chosen optical surface roughness values $g$. The simulation results of virtual measurements of strongly scattering to ideally reflecting surfaces (in Fig. \ref{fig:2_mts_logo_parameter_variation_images}, becoming optically smoother from top to bottom) differ greatly from each other, which is in accordance with experienced measurement behavior.
In this example, the irregular structured and at the same time strongly optically scattering surface of the virtual measuring sample reflects a large portion of the illumination light back towards the objective and thus onto the detector (see Fig. \ref{fig:2_mts_logo_parameter_variation_images}, upper row, $g=0.3$). In contrast, when the irregular structured but optically ideal smooth surface is measured virtually (see Fig. \ref{fig:2_mts_logo_parameter_variation_images}, bottom row, $g=1$), only a small amount of light reaches the detector because most of the illumination light is reflected past the objective due to the high gradients of the sample's surface. Thus, consideration of optical surface roughness, e.g. in the format of a scattering function, is an important model component for a realistic virtual measurement and usually cannot be neglected.

\subsection{Comparison of real and virtual imaging}
\begin{figure} [ht]
   \begin{center}
   \begin{tabular}{c} 
   \includegraphics[width=17.15cm]{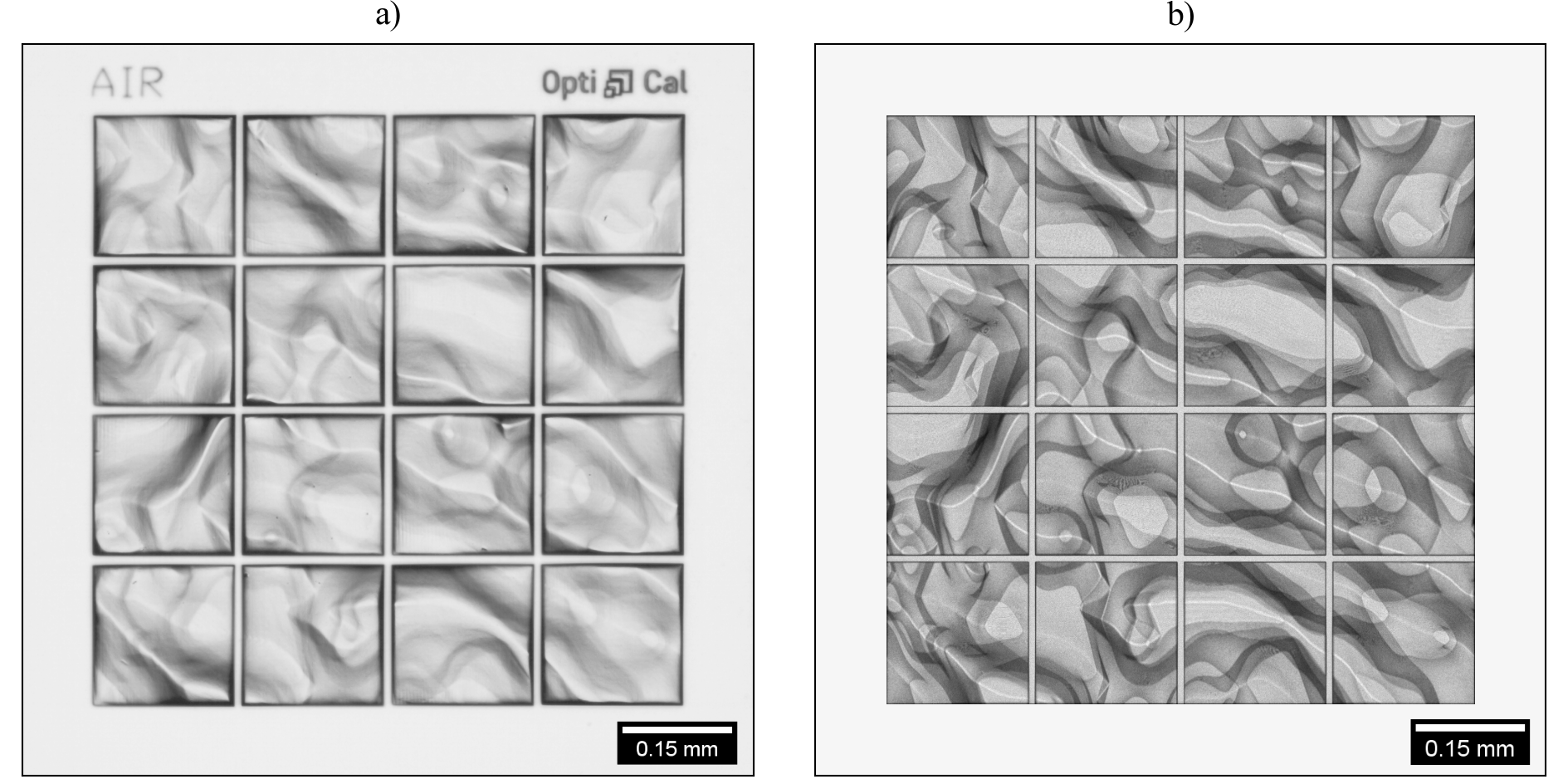}
   \end{tabular}
   \end{center}
   \caption[example] 
   { \label{fig:3_Comparison_AIR_real_virtual} 
Comparison of a) real and b) virtually acquired detector images when measuring an areal irregular roughness (AIR) reference surface. An optical surface roughness of $g=0.8$ was assigned to the virtual deterministic surface and the virtual detector image was generated by tracing $N_{\text{Rays}}=1E10$ light rays.}
   \end{figure}

To demonstrate the fidelity of the comprehensive virtual measurement process, real and virtual detector images of a reference surface are qualitatively compared with each other. The underlying reference surface is an areal irregular roughness calibration geometry (type AIR) with a deterministic structure based on a real engineering surface. This structure is one of six calibration structures found on the Opti-Cal universal standard. The Opti-Cal standard is a calibration specimen capable of providing the full set of relevant metrological characteristics on a single sample \cite{Eifler.2018}.
In Fig. \ref{fig:3_Comparison_AIR_real_virtual}, real (Fig. \ref{fig:3_Comparison_AIR_real_virtual}a) and virtually generated (Fig. \ref{fig:3_Comparison_AIR_real_virtual}b) detector images of the previously described reference surface are compared with each other. The rasterization of the reference surface into 4$\times$4 sub-surfaces, as shown in Fig. \ref{fig:3_Comparison_AIR_real_virtual}, has a manufacturing-related background. Optical topography measuring instruments with low magnification (e.g. 5X) provide measuring fields of $\sim 1~\mathrm{mm}~\times 1~\mathrm{mm}$. When calibrating these instruments, fields of view of this size must be taken into account. However, since the operating range of the direct laser writing process used for manufacturing is limited to $\sim 0.3~\mathrm{mm}~\times 0.3~\mathrm{mm}$, structures of this size have to be subdivided into sub-surfaces \cite{Eifler.2018}. 

It is important to note that the optical system of the FV instrument used to acquire the real detector image is not the same as the optical system of the virtually modeled instrument. However, the numerical aperture ($NA$) of the lenses of both systems (real and virtual) are chosen to be the same, so the results are at least qualitatively comparable. For the virtual measurement, an optical surface roughness of $g=0.8$ was assigned to the deterministic reference surface and $N_{\text{Rays}}=1E10$ rays were used to generate the virtual detector image.

A qualitative comparison of the captured (Fig. \ref{fig:3_Comparison_AIR_real_virtual}a) and virtually generated (Fig. \ref{fig:3_Comparison_AIR_real_virtual}b) intensity distributions imaged on the detector confirms the potential accuracy of the virtual measurement. In this example only the contrast ratios between bright and dark areas are slightly more pronounced in the virtual image in this example. It should be noted that no diffraction effects have been considered in the virtual measurement so far. Including diffraction effects would further reduce the resulting contrasts and thus further improve the fidelity of the virtual image in this example. A simple way to include diffraction effects would be to apply a theoretical or measured point spread function of the measuring device to the virtual detector image.

\subsection{Virtual areal topography measurement}
Finally, the evaluability of virtually generated image stacks is demonstrated. For this purpose, an exemplary image stack of the already presented AIR reference surface, which was generated by applying the insights from the simulations in connection with sections 4.1 and 4.2, is evaluated using the software MountainsMap\textregistered~and its integrated function ``3D reconstruction from multifocus images". The result of this evaluation is an areal surface topography reconstructed on the basis of the virtual measurement. The simulation parameters that were defined for generating the image stack are summarized in Tab. \ref{tab:Topography_Simulation_Parameters}.

\begin{table}[ht]
\caption{Simulation parameters to generate the image stack of the virtual areal surface topography measurement.} 
\label{tab:Topography_Simulation_Parameters}
\begin{center}       
\begin{tabular}{|l|l|} 
\hline
\rule[-1ex]{0pt}{3.5ex}  Number of light rays traced per image $N_{\text{Rays}}$  & 1E10 \\
\hline
\rule[-1ex]{0pt}{3.5ex} Digital detector resolution $\Delta x$ & $\approx 1~\mathrm{\upmu m}$ \\
\hline
\rule[-1ex]{0pt}{3.5ex} Axial spacing between individual detector images $\Delta z$ & 0.05 $\mathrm{\upmu m}$\\
\hline
\rule[-1ex]{0pt}{3.5ex} Optical surface roughness $g$ & 0.8 \\
\hline
\rule[-1ex]{0pt}{3.5ex} Averaged simulation time per image $t_{\text{Image}}$ & $\sim$ 660 s\\
\hline

\end{tabular}
\end{center}
\end{table}

\begin{figure} [ht]
   \begin{center}
   \begin{tabular}{c} 
   \includegraphics[width=17.15cm]{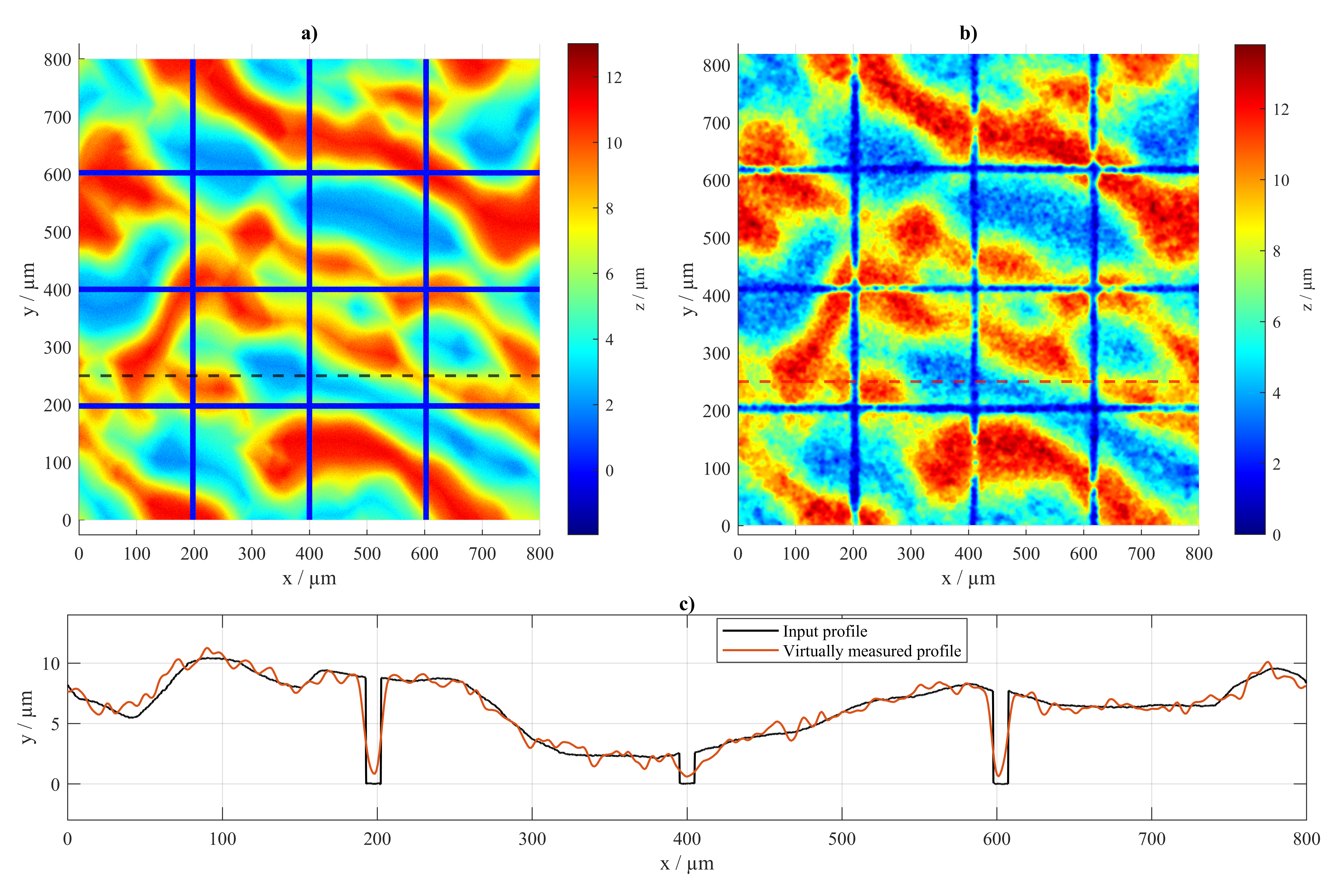}
   \end{tabular}
   \end{center}
   \caption[example] 
   { \label{fig:4_Topography_Comparison} a) Input surface topography of areal irregular roughness (AIR) reference sample; b) Virtually measured and evaluated AIR surface topography; c) Exemplary profiled comparison of initial and virtually measured AIR surface topographies imaged in a) and b). 
}
   \end{figure}

The aforementioned input topography of the reference surface was additionally superimposed with a micro-roughness which was generated by adding Gaussian noise with $(\mu_{\text{Noise}},\sigma_{\text{Noise}})=(0,0.1)~\mathrm{\upmu m}$, filtered with a Gaussian filter ($\sigma_{\text{Kernel}}=1~\mathrm{\upmu m}$).
The resulting surface topography, along with the input topography used, is shown in Fig. \ref{fig:4_Topography_Comparison}, both as areal surface and as exemplary profile.
   
A qualitative comparison of the two surfaces shown in Fig. \ref{fig:4_Topography_Comparison}a and Fig. \ref{fig:4_Topography_Comparison}b leads to the conclusion that the shape of the surface to be measured can be reconstructed given the virtual measurement conditions defined in this example. At the same time, it can be observed that the reconstructed surface topography deviates strongly from the underlying input surface with respect to its micro structure. A robust evaluation of the roughness is not possible given the entire set of conditions present in this example (measuring device hardware, optical and topographic properties of the measured sample, defined measuring process parameters, applied evaluation algorithm). However, this result could serve as a starting point of a parameter study to identify optimization potentials of the measuring device and/or adjustable measuring process parameters.

\section{Conclusions}
This paper presents a comprehensive approach to model optical measuring systems utilizing the NVIDIA OptiX ray tracing engine, using a virtual focus variation instrument for measuring areal surface topographies as an example. The key aspects that make OptiX attractive for scientific modeling are its GPU-accelerated parallel computing capabilities, access to powerful ray tracing algorithms, and a high level of flexibility in model design.
The virtual focus variation instrument developed in this paper demonstrates its ability to virtually reproduce accurate intensity distributions of a realistic scene in a reasonable time. An exemplary comparison of real and virtually reproduced images of an engineering surface further validates the fidelity of the approach. Furthermore, the evaluability of virtual measurements is generally demonstrated by reconstructing an exemplary areal surface topography based on a virtually generated stack of focus variation microscopy images.
Overall, the comprehensive approach presented here provides a solid foundation for the application of virtual measuring techniques improving the overall process of designing optical systems, optimizing measuring process parameters, and testing signal processing algorithms. 
In future research, additional optical effects such as diffraction, coherence, and polarization effects will be integrated into the underlying optical models, allowing the simulation of further optical measuring systems, such as white light interferometers, and thus extending the applicability of virtual metrology to a wider range of scientific and engineering contexts.

\acknowledgments
We thank Bruker Alicona for providing the virtual optical system used to perform the virtual measurements.


The 20IND07 TracOptic project has received funding from the EMPIR programme co-financed by the Participating States and from the European Union’s Horizon 2020 research and innovation programme.
\begin{figure}[h!]
\centering
 \includegraphics[width=8cm]{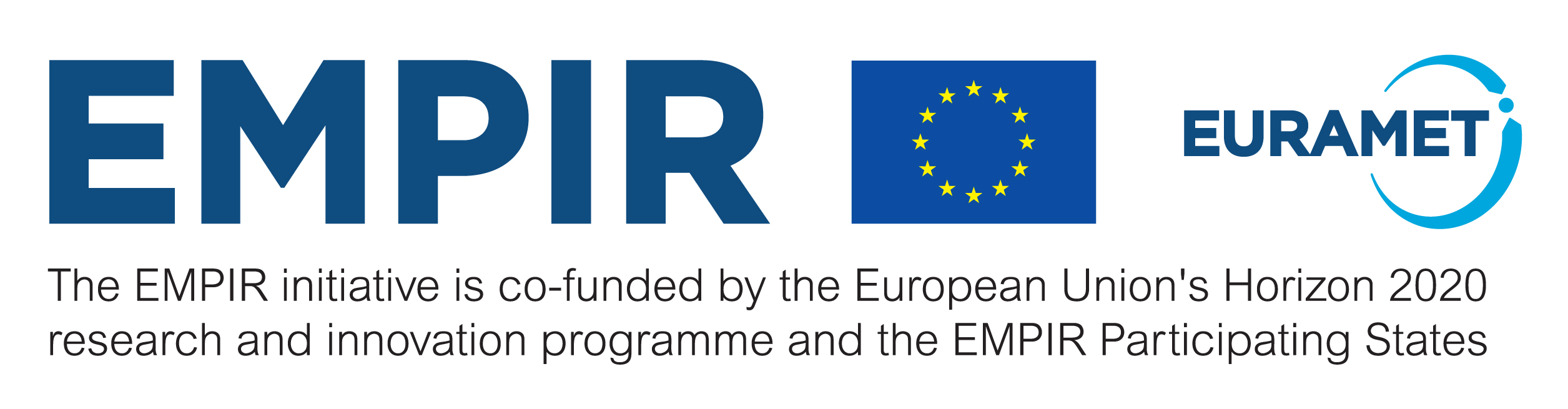}
 
\end{figure}

\bibliography{Literatur_SPIE_Tracoptic} 
\bibliographystyle{spiebib} 

\end{document}